\documentstyle[12pt]{article}

\newlength{\extraspace}
\newlength{\extraspaces}
\begin{document}

\addtolength{\baselineskip}{.8mm}

\thispagestyle{empty}

\begin{flushright}
{\sc FERMILAB-PUB}-97/046-T\\
  hep-th/9703009\\
\hfill{February 28, 1997  }\\ 
\end{flushright}
\vspace{.3cm}

\begin{center}
{\large\sc{ S-DUALITY AND COMPACTIFICATION OF TYPE IIB SUPERSTRING ACTION
}}\\[15mm]

{\sc Jnanadeva Maharana\footnote{E-mail: maharana@fnth29.fnal.gov}}\\ 
[2mm]
{\it Fermi National Accelerator Laboratory\\
P. O. BOX 500\\
Batavia, Illinois 60510\\
and\\
Institute of Physics\\
Bhubaneswar 751005\\
India} \\[15mm]


{\sc Abstract}

\begin{center}
\begin{minipage}{14cm}

The $\bf {SL(2,R)}$ invaraint ten dimensional type IIB superstring effective 
action is
compactified on a torus to D spacetime dimensions. The transformation
properties of scalar, vector and tensor fields, appearing after the dimensional 
reduction, are obtained in order to
maintain the $\bf {SL(2,R)}$ invariance of the reduced effective action.
The symmetry of the action enables one to generate new string vacua from
known configurations. As illustrative examples, new black hole solutions
are obtained  in five and four dimensions from a given set of solutions of
the equations of motion.

\end{minipage}
\end{center}

\end{center}

\noindent

\vfill
\newpage
\pagestyle{plain}
\setcounter{page}{1}
\stepcounter{subsection}
\renewcommand{\footnotesize}{\small}
\def\be{\begin{equation}}
\def\ee{\end{equation}}
\def\bea{\begin{eqnarray}}
\def\eea{\end{eqnarray}}

It is recognised that dualities play a central role in our understanding of
the dyanamics of string theory[1]. The intimate connections between various
superstring theories and the nonperturbative features of these theories
in diverse dimensions are unravelled by the web of duality relations[2,3]. 
The
S-duality transformation relates strong and weak coupling phases of a given
theory in some cases, whereas in some other situation strong and weak coupling
regimes of two different theories are connected. One familiar example is the
heterotic string toroidally compactified from ten to four dimension and for
such a theory S-duality is the generalization of the familiar 
electric-magnetic duality. Another situation arises in six spacetime
dimensions; when the ten dimensional  
hetetoric string is compactified on $T^4$. The  
S-duality, on this occasion, 
relates the six dimensional heterotic string to type IIA theory compactified on $K_3$.

It was conjectured that type IIB theory in ten dimensions is endowed with
$\bf {SL(2,Z)}$ symmetry [4,5]. There is mounting evidence for this symmetry  
and it has played a very
important role in  providing deeper insight into
the nonperturbative attributes  of type IIB theory [6,7]. Furthermore, 
there is an intimate connection between type IIB theory compactified on
a circle and the M-theory compactified on $T^2$
 leading to a host of interesting
results [8]. We recall that the bosonic massless 
excitations of type IIB theory
consist of graviton, dilaton and antisymmetric  tensor  in the NS-NS sector;
denoted by $\hat G_{MN}, \hat \phi$ and $ \hat B^{(1)}_{MN}$, respectively.
The R-R counterparts are 'axion', $\hat \chi$, another antisymmetric tensor
field, $\hat B^{(2)}_{MN}$ and a four index antisymmetric potential, ${\hat C}
_{MNPR}$, 
with self-dual field strength. The Lorentz indices in ten dimensions are
denoted by M,N,P,... and the field are defined with a hat. The complex moduli,
$\hat \lambda = \hat \chi + ie^{-\hat \phi} $ is known to transform nontrivially
under  $\bf {SL(2,R)}$ and same is the case for the 
two second rank tensor fields $\hat B^(1)$ and $B^{(2)}$
 (see below for the exact 
transformation rules). The $\bf {SL(2,R)}$ eventually breaks to the robust 
discrete
symmetry  $\bf {SL(2,Z)}$.

The purpose of this investigation is to study toroidal compactification of
the type IIB theory and implications of $\bf {SL(2,R)}$ symmetry for the
reduced action. Furthermore, we would like know, what are the types
of interaction terms involving  the moduli, dilaton and axion, 
 are permitted 
when we impose $\bf {SL(2,R)}$ symmetry. It will be shown that a manifestly
$\bf {SL(2,R)}$ invariant reduced action can be written down by defining 
the transformation properties for the scalar, gauge and tensor  fields, 
  which appear as a consequence of  
 toroidal compactification of the ten dimensional theory. We shall show that
the interactions involving the complex moduli, expressed in terms of the 
dilaton and the axion, can be restricted by demanding the $\bf {SL(2,R)}$
invariance of the effective action.

The compactifications of type IIA and type IIB theories 
as we go from ten to nine dimensions have been 
studied by Bergshoeff, Hull and Otin [9] and they have explored 
implications of  various dualities for this compactification; more recently,
Andrianopoli and collaborators [10] have studied compactification of type II
theories and M-theory in various dimensions.
 It is well known that type IIA and type IIB theories
are related by T-duality below ten dimensions [11]. 
Furthermore, in lower dimensions
the S-duality combines with the T-duality leading to U-duality; for example
in 8-dimensions, the U-duality group is $\bf {SL(3,Z)} \times \bf {SL(2,Z)}$ and
various  
branes belong to representations [12] of this larger group. Recently, the
five dimensional string effective action, obtained by toroidal compactification
of type IIB superstring action, has attracted considerable of attention
in establishing  Beckenstein-Hawking area-entropy relations for extremal
black holes and the near extremal ones [13-18]. Therefore, it is of interest
to obtain type IIB effective action, through dimensional reduction, in 
lower dimensional spacetime and explore the implications of $\bf {SL(2,R)}$ 
duality transformations. 
  
Let us consider the ten dimensional action for the type IIB theory:

\bea \hat S& =& {1\over {2 {\kappa^2}}} \int d^{10}x \sqrt{-\hat G} \bigg\{
e^{-2\hat \phi}\left ( \hat R + 4(\partial \hat \phi)^2 - {1\over 12} 
\hat H^{(1)}_
{MNP} \hat H^{(1) MNP}\right) - {1\over 2} (\partial \hat \chi)^2\nonumber\\ &
&  
- {1\over 12}
{\hat \chi}^2 \hat H^{(1)}_{MNP} \hat H^{(1) MNP} 
- {1\over 6} \hat \chi
\hat H^{(1)}_{MNP} \hat H^{(2)MNP} -{1\over 12} \hat H^{(2)}_{MNP} \hat
H^{(2) MNP} \bigg\} \eea

Here $\hat G_{MN}$ is the ten dimensional metric in the string frame and 
$\hat H^{(1)}$ and $\hat H^{(2)}$ are the field strengths associated with the
potentials $\hat B^{(1)}$ and $\hat B^{(2)}$ respectively.  It is well known
that in ten dimensions, it is not possible to construct a covariant action [19]
 for ${\hat C}_{MNPR}$ with self-dual field strength and therefore, we have
set this field to zero throughout this paper; however, one can 
dimensionally reduce this field while carrying out compactification; we
set it to zero for convenience. In order to
express the action in a manifestly  $\bf {SL(2,R)}$ invariant form [20,7], 
recall that the axion and the dilaton parametrize [21] 
the coset ${\bf {SL(2,R)}}
\over {SO(2)}$. We
over to the Einstein frame through the conformal transformation $\hat g_{MN} =
e^{-{1\over 2} \hat \phi} \hat G_{MN}$ and the action (1) takes the form

\bea \hat S_E = {1\over {2{\kappa ^2}}} \int d^{10} x \sqrt {-\hat g} 
\bigg\{  
\hat {R}_{\hat g} 
+ {1\over 4} Tr (\partial _N \hat { \cal  M} \partial ^N \hat {\cal M}^{-1})
- {1\over 12} \hat {\bf { H}}^T_{MNP} \hat {\cal M} \hat {\bf { H}}^{MNP}
 \bigg\} \eea

Here $\hat {R}_{\hat g}$ is the scalar curvature computed from the Einstein 
metric. The
matrix $\hat {\cal M}$ is defined as:
\bea \hat {\cal M} = \pmatrix {\hat {\chi}^2 e^{\hat \phi} + e^{- \hat \phi} &
\hat \chi e^{\hat \phi}\cr
\hat \chi e^{\hat \phi} & e^{\hat \phi}\cr},
\qquad \bf \hat H = \pmatrix {\hat
H^{(1)}\cr \hat H^{(2)}\cr}  \eea

Note that $det \hat {\cal M} $ is unity. The action is invariant under following
transformations,
\bea \hat { \cal M} \rightarrow \Lambda \hat {\cal M} {\Lambda }^T, \qquad H
 \rightarrow ({\Lambda }^T)^{-1} H, \qquad \hat { g}_{MN} \rightarrow \hat 
 {g}_{MN}, 
           \eea

where $\Lambda \in \bf {SL(2,R)}$. Let us introduce a $2 \times 2$ matrix,
$\Sigma$ and consider  a generic form of $\Lambda$

\bea \Lambda = \pmatrix {a & b\cr c & d\cr}, \qquad \Sigma = \pmatrix {0 & i\cr
-i & 0\cr} \eea
with $ad - bc = 1$. It is easy to check that, 
\bea \Lambda \Sigma {\Lambda }^T = \Sigma, \qquad \Sigma \Lambda \Sigma 
= {\Lambda }^{-1} \eea
 and
\bea {\hat {\cal M}} \Sigma {\hat {\cal M}} = \Sigma , 
\qquad \Sigma {\hat {\cal M}} {\Sigma} = {\hat {\cal M}} ^{-1}  \eea

Thus $\Sigma $ plays the role of $\bf {SL(2,R)}$ metric and the symmetric 
matrix
     $ \hat {\cal M} \in \bf {SL(2,R)}$.

It is evident that the second term of eq(2) can be written as
\bea {1\over 4} Tr[  {\partial} _N {\hat {\cal M}} \Sigma 
{\partial} ^N {\hat {\cal M}}
\Sigma ] \eea

The Einstein equation can be derived by  varying the action with respect to
the metric and the equation of motion associated with the antisymmetric
tensor fields can be obtained in a straight forward manner. The 
$\hat {\cal M}$ -equation of motion follows from the variation of the action if
we keep in mind that $\hat {\cal M}$ is a symmetric $\bf {SL(2,R)}$ matrix
satisfying the properties mentioned above. Thus, if we consider an 
infinitesimal transformation, we arrive at following relations.
\bea \Lambda = {\bf 1} + \epsilon , \qquad \Lambda \in {\bf {SL(2,R)}} \eea
\bea \epsilon \Sigma + \Sigma {\epsilon }^T = 0 , \qquad {\hat {\cal M}} 
\rightarrow \epsilon {\hat {\cal M}} + {\hat {\cal M}}  {\epsilon} ^T 
+ {\cal M} \eea

Now the desired equation of motion, derived from the above action, is 
\bea {\partial} _M (\sqrt {-\hat g} {\hat {g}^{MN}} {\hat {\cal M}} \Sigma
{\partial} _N {\hat {\cal M}} \Sigma ) - {1\over 6} {\hat {\bf H}}^T 
{\hat {\cal M}}
 {\hat {\bf H}}  = 0 \eea
Note that this is a matrix equation of motion and we have suppressed the 
indices for notational conveniences. It is worthwhile, at this stage to 
point out some similarities with the the global $O(d,d)$ symmetry that arises
 when one considers toroidal compactifications to lower spacetime dimensiona.  
The
metric $\Sigma$ is analogous to the metric, $\eta $, associated with the
$O(d,d)$ transformations and the $\hat {\cal M}$ equation of motion bears
resemblance with the corresponding $M$-matrix of the $O(d,d)$ case [22,23].

Let us consider compactification of the action of the type IIB theory on
a d-dimensional torus to obtain the dimensionally reduced effective action
[24-27]. 
Let the D-dimensional spacetime coordinates be denoted
by $x^{\mu}, \mu = 1, 2, .. D-1$ and the internal coordinates be labelled by
$y^{\alpha}$ and $\alpha$ takes d values so that $D+d=10$. The following choice
of 10-dimensional vielbein [24,25] is convenient to derive the reduced action:
\bea {\hat { e}^{A}_{M}} = \pmatrix { e^{r}_{\mu} & {\cal A}^{\beta}_{\mu}
E^{a}_{\beta}\cr
0 & E^{a}_{\alpha}\cr } \eea
The D-dimensional spacetime metric is $g_{\mu \nu} = e^{r}_{\mu}e^{s}_{\nu}
\eta _{rs}$, $\eta _{rs}$ being  D-dimensional Lorentzian signature flat metric
 and the internal metric is 
${\cal {G}}_{\alpha \beta} = E^{a}_{\alpha}
E^{b}_{\beta} \delta_{ab}$. In our notation above,
 $A, r$ and $a$ denote the local Lorentz
indices of $\hat {e}^{A}_{M}, e^{r}_{\mu}$ and $E^{a}_{\alpha}$ respectively
and $M, \mu$ and $\alpha$ are the corresponding global indices. 
Let us  assume that the backgrounds are
independent of the set of internal coordinates $y^{\alpha}$ and derive the 
reduced effective action. 
Note that, with
choice of our vielbein, $\sqrt {-\hat g} = \sqrt {-g} \sqrt {\cal G}$. The
10-dimensional Einstein-Hilbert Lagrangian density is decomposed,
when we go down to lower dimensions, as  sum of
three terms consiting of ${\sqrt{-g}} R$, kinetic energy term for the scalars,
$\cal {G} _{\alpha \beta} $ and the kinetic energy term for the Abelian gauge
fields $\cal {A}^{\alpha}_{\mu}$. The matrix $\hat {\cal M}$ defined in terms
of the dilaton and axion becomes a matrix which carries $x$-dependence only and
we denote it by $\cal M$ from now on.

The term ${\hat {\bf H}}^T {\hat {\cal M}} {\hat {\bf H}}$ is expressed as
sum of three terms (each of the term is a scalar with no free index): 
a term with one Lorentz index and two internal indices, another term which
has two Lorentz indices and one internal index and a term with three Lorentz
indices. The term with all internal indices, $H_{\alpha \beta \gamma} 
$,  vanishes, since we assume 
$y$-indepence of backgrounds and $\bf H$ always involves derivatives. When we 
want to obtain a D-dimenional tensor from a given 10-dimensional one, we
first convert the global indices of the 10-dimensional tensor to local
indices by multiplying suitable numbers of $\hat e$'s and ${\hat e}^{-1}$'s
in ten dimensions, then we multiply with $e$'s and $e^{-1}$ of D-dimensions.
For the case at hand,
\bea H^{(i)}_{\mu \alpha \beta} = \partial _{\mu} B^{(i)}_{\alpha \beta} \eea
\bea H^{(i)}_{\mu \nu \alpha} = F^{(i)}_{\mu \nu \alpha} - B^{(i)}_{\alpha 
\beta} {\cal F}^{\beta}_{\mu \nu} \eea 
Where, the upper index $i=1,2$,  
${\cal F}^{\alpha}_{\mu \nu} = \partial _{\mu}{\cal A}^{\alpha}_{\nu} -
\partial _{\nu} {\cal A}^{\alpha}_{\mu}$ and $F^{(i)}_{\mu \nu \alpha} =
\partial _{\mu} A^{(i)}_{\nu \alpha} - \partial _{\nu} A^{(i)}_{\mu \alpha}$
and the gauge potential $A^{(i)}_{\mu \alpha} = {\hat B}^{(i)}_{\mu \alpha} +
B^{(i)}_{\alpha \beta} {\cal A}^{\beta}_{\mu}$. 
 The  antisymmetric
field strength in D-dimensions takes the following form. 
\bea H^{(i)}_{\mu \nu \rho} = \partial _{\mu}B^{(i)}_{\nu \rho} - {1\over 2}[
{\cal A}^{\alpha}_{\mu}F^{(i)}_{\nu \rho \alpha} + A^{(i)}_{\mu \alpha}{\cal
F}^{\alpha}_{\nu \rho}] + cycl. perm \eea
We mention in passing,  the presence of Abelian Chern-Simons term in the
expression for the field strength [25], resulting from the dimensional 
reduction. The two form potential is defined as
\bea B^{(i)}_{\mu \nu} = {\hat B}^{(i)}_{\mu \nu} + {1\over 2} {\cal A}^{
\alpha}_{\mu}A^{(i)}_{\nu \alpha} - {1\over 2}{\cal A}^{\alpha}_{\nu}A^{(i)}_
{\mu \alpha} - {\cal A}^{\alpha}_{\mu} B^{(i)}_{\alpha \beta}{\cal A}^{\beta}_
{\nu} \eea
Notice that ${\bf H}_{\mu \nu \rho}$ is antisymmetric in all its tensor 
indices as
should be the case. 

The 10-dimensional effective action is invariant under general
coordinate transformations as well as the gauge transformations associated
with the two antisymmetric tensor fields. When we examine the local symmetries 
of the theory in D-dimensions after dimensional reduction, we find that there
is general coordinate transformation invariance in D-dimensions. The
Abelian gauge transformation, associated with ${\cal A}^{\alpha}_{\mu}$,
has its origin in 10-dimensional general coordinate transformations. The
field strength $H^{(i)}_{\mu \nu \alpha}$ is invariant under a suitable gauge
transformation once we define the gauge transformation for $F^{(i)}_{\mu \nu 
\alpha}$ since ${\cal F}^{\alpha}_{\mu \nu}$ is gauge invariant under the
gauge transformation of ${\cal A}$-gauge fields. Finally, the tensor field
strength  $H^{(i)}_{\mu \nu \rho}$, defined above, can be shown to be gauge
invariant by defining appropriate gauge transformations for
 $B^{(i)}_{\mu \nu}$; $\delta B^{(i)}_{\mu \nu} = \partial_{\mu}\xi^{(i)}_{\nu}
- \partial_{\nu}\xi^{(i)}_{\mu}$. 

\newpage

The D-dimensional effective action takes the following form  
\bea S_E & =& \int d^D x {\sqrt {-g}}{\sqrt {\cal G}} \bigg\{ R + {1\over 4}[
{\partial}_{\mu}{\cal G}_{\alpha \beta} {\partial}^{\mu}{\cal G}^{\alpha \beta}
+ {g}^{\mu \nu} {\partial}_{\mu}{ln \cal G}
{\partial}_{\nu}{ ln \cal G} - g^{\mu \lambda}
g^{\nu \rho} {\cal G}_{\alpha \beta}{\cal F}^{\alpha}_{\mu \nu}{\cal F}^{\beta}
_{\lambda \rho}]\nonumber\\ & & -{1\over 4} {\cal G}^{\alpha \beta}{\cal G}^{
\gamma \delta} {\partial}_{\mu}B^{(i)}_{\alpha \gamma}{\cal M}_{ij}{\partial}^{
\mu} B^{(j)}_{\beta \delta} - {1\over 4} {\cal G}^{\alpha \beta}g^{\mu \lambda}
g^{\nu \rho} H^{(i)}_{\mu \nu \alpha}{\cal M}_{ij} 
H^{(j)}_{\lambda \rho \beta}\nonumber\\ & & -{1\over {12}} H^{(i)}_{\mu \nu 
\rho} {\cal M}_{ij} H^{(j) \mu \nu \rho} + {1\over 4} Tr ({\partial}_{\mu}
{\cal M} \Sigma {\partial}^{\mu}{\cal M} \Sigma) \bigg\} \eea
 
The above action is expressed in the Einstein frame, $\cal G$ being determinant
of ${\cal G}_{\alpha \beta}$.
 If we demand 
  $\bf {SL(2,R)}$ invariance of the above action, then  the backgrounds
are required to  satisfy following transformation properties:
\bea {\cal M} \rightarrow \Lambda {\cal M}{\Lambda}^T , \qquad H^{(i)}_{\mu \nu
\rho} \rightarrow {( {\Lambda}^T)^{-1}}_{ij}H^{(j)}_{\mu \nu \rho} \eea
\bea A^{(i)}_{\mu \alpha} \rightarrow {({\Lambda}^T)^{-1}}_{ij}A^{(j)}_{\mu 
\alpha}, \qquad B^{(i)}_{\alpha \beta} \rightarrow {({\Lambda}^T)^{-1}}_{ij}
B^{(j)}_{\alpha \beta} \eea
and
\bea g_{\mu \nu} \rightarrow g_{\mu \nu}, \qquad {\cal A}^{\alpha}_{\mu} 
\rightarrow {\cal A}^{\alpha}_{\mu}, \qquad {\cal G}_{\alpha \beta}
\rightarrow {\cal G}_{\alpha \beta} \eea
and $\Lambda \in {\bf {SL(2,R)}}$.

It is evident from the D-dimensional action that dilaton and axion interact
with antisymmetric tensor fields, gauge fields and the scalars 
due to the presence of $\cal M$ matrix in various terms and these 
interaction terms respect
the ${\bf SL(2,R)}$ symmetry. 
It is important know what type of dilatonic potential is admissible in 
the above action which respects the S-duality symmetry. The only 
permissible  interaction terms, preserving the symmetry, are of the form
\bea Tr{[ {\cal M} \Sigma]}^n , \qquad {n \in {\bf Z}} \eea
It is easy to check using the properties of $\Sigma$ and ${\cal M}$ matrices;
such as $Tr ({\cal M} \Sigma) = 0 $ and $Tr ({\cal M}\Sigma {\cal M}\Sigma)
= 2 $,
that
\bea Tr {[{\cal M} \Sigma]}^n = 0, \qquad and , 
\qquad Tr {[{\cal M} \Sigma]}^n = 2,  \eea
For odd $n \in {\bf Z}$ and even $n \in {\bf Z}$ respectively.
Therefore, we reach a surprizing conclusion that  the presence of 
interaction terms of the
form in eq.(21)only adds constant term which amounts to adding 
cosmological constant term to the
reduced action. Note that the Einstein metric is $\bf {SL(2,R)}$ invariant
and one can add terms involving higher powers of curvature (higher derivatives
of metric) to the action and maintain the symmetry. However, we are considering
the case when the gravitational part of the action  has the 
Einstein-Hilbert term only.

Now we proceed to  present illustrative examples which demonstrate the 
application  of
solution generating technique to derive new backgroundis by implementing
${\bf SL(2,R)}$ transformations on 
 an initial set which satisfy the equation of motion. The first example
is a five dimensional effective action [28] which has the  following form
\bea \int d^5 x {\sqrt {-g}} \bigg\{ R + {1 \over 4} Tr ({\partial}_{\mu}
{\cal M} \Sigma {\partial}^{\mu}{\cal M} \Sigma) - {1\over {12}} e^{-\phi}
H^{(1)}_{\mu \nu \rho} H^{(1)\mu \nu \rho} - {1\over 4} e^{\phi} F_{\mu \nu}
F^{\mu \nu} \bigg\} \eea
This action corresponds to the following  choice of backgrounds   
\bea H^{(2)}_{\mu \nu \rho} = 0, \qquad {\cal A}^{\alpha}_{\mu} = 0, \qquad
H^{(1)}_{\mu \nu \alpha} = 0, \qquad \chi = 0, \qquad  
  B^{(i)}_{\alpha \beta} = 0, \qquad {\cal G} = 1 \eea
Of the five field strengths, $F_{\mu \nu \alpha}$, $\alpha = 5, 6, 7, 8, 9$,
coming from compactification of $H^{(2)}_{MNP}$, 
we choose only one of them to be nonvanishing  and set rest to zero. Moreover,
${\cal M} =$ diag $($ $e^{-\phi}$, $e^{\phi}$ $)$, since $\chi=0$. This five
dimensional action is quite similar to the one considered by Strominger
and Vafa [13] in their seminal paper in which they derived the 
Beckenstein-Hawking
area-entropy relation for a class of five dimensional extremal black holes.
It is easy to see that there will be conserved charges $Q_H$ and $Q_F$
proportional to
\bea {\int }_{S^3} *e^{-\phi} H^{(1)}, \qquad {\int}_{S^3} *e^{\phi} F \eea
respectively. The dilaton equation takes the form 
\bea (\nabla \phi)^2 + {1\over {12}} e^{-\phi} H^{(1)}_{\mu \nu \rho}
H^{(1) \mu \nu \rho} - {1\over 4}e^{\phi} F_{\mu \nu}F^{\mu \nu} = 0 \eea
Thus if we have constant dilaton configuration $\phi _{c}$, then $e^{\phi_c}$
will be proporational to $({Q_{F}\over {Q_H}})^2$. Since the action is
invariant under ${\bf SL(2,R)}$ transformation, we can generate new backgrounds
with nonzero $H^{(2)}$,  $F^{(1)}$ and $\chi$, by suitable implementaion of the
symmetry transformation [29,30]; 
note that the $F$ appearing in the action is $F^{(2)}$ 
according to our choice. The simplest form of $\Lambda = \pmatrix {0 & -1\cr
1 & 0\cr } $ takes $e^{\phi}$ to $e^{-\phi}$ and is just the strong-weak duality
transformation. A more interesting transformation  is when 
\bea \Lambda = {1\over{\sqrt 2}} 
\pmatrix {cosh{\theta} - sinh{\theta} & cosh{\psi} + 
sinh{\psi}\cr sinh{\psi} - cosh{\psi} & cosh{\theta} + cosh{\theta}\cr }\eea
Here $\theta$ and $\psi$ are two 'boost' parameters. It is easy to construct
the new matrix ${\cal M}'$ from eq.(18) and we find 
\bea e^{{\phi}'} = {1\over 2} 
[e^{-\phi}(cosh{\psi}+sinh{\psi})^2 + e^{\phi}(cosh{\theta}
+ sinh{\theta})^2]  \eea
and
\bea {\chi}'e^{{\phi}'}& =& {1\over 2}[e^{-\phi}(cosh{\theta} - 
sinh{\theta})(
cosh{\psi} + sinh {\psi})\nonumber\\ & &  + 
e^{\phi}(cosh{\theta} + sinh{\theta})(sinh{\psi}
- cosh{\psi})] \eea 
Now we have both the gauge field strengths and they are given by  
\bea F^{(1)}_{\mu \nu} = -{1\over{\sqrt 2}} (cosh {\psi} + sinh{\psi})F_{\mu 
\nu} \eea
\bea F^{(2)}_{\mu \nu} = {1\over{\sqrt 2}}(cosh{\theta} - sinh{\theta})F_{\mu
\nu} \eea 
where  $F_{\mu \nu}$ appearing in the right  hand side of the above equation
is the one that was introduced  in the five dimensional action, eq.(23). 
Furthermore, the new antisymmetric tensor field strengths are,
\bea H^{(1)'}_{\mu\nu\rho} = {1\over{\sqrt 2}}(cosh {\theta} + sinh {\theta})
H^{(1)}_{\mu \nu \rho} \eea
\bea H^{(2)'}_{\mu\nu\rho} = {1\over{\sqrt 2}}(cosh {\theta} - sinh {\theta})
H^{(1)}_{\mu\nu\rho} \eea
We
note that the Einstein metric is invariant under these transformations
and therefore, the spacetime geometry remains unchanged and we expect that the
Hawking temperature, $T_H$ to be  the same for the  family of black holes
obtained through this procedure. Indeed, they will carry charges with
respect to the two field strengths as well as have axionic charges. Note that
these charges will be characterised by the pair of parameters $\theta$ and
$\psi$. 
Of course,
the residual unbroken symmetry is $\bf {SL(2,Z)}$ and then the transformation
matrix $\Lambda$ will have integer elements, satisfying the requisite
constraints.

Our next example is the four dimensional black hole solution discussed by
Shapere, Trivedi and Wilczek [31]. Let us recall that the charged black hole
solution was obtained from  an effective action which had metric,
dilaton and a  gauge field. Next, these authors obtained solutions in
the presence of the axion. The axion field appears after one implements
Poincare duality transformation on the field strength of the antisymmetric
tensor (in their case it came from the NS-NS sector).

If we start with  an action with graviton, dilaton and a gauge field, we can 
generate new solutions in the following ways.

{\it (i)} We can envisage  the gauge field arising from the compactification
of the metric, say one of the ${\cal A}^{\alpha}_{\mu}$ fields. Then,
under the $\bf {SL(2,R}$ transformations $\cal A$ remains invariant according
to our rules; however, we shall generate nontrivial $\chi$ field which comes
from the RR sector and the transformed dilaton will be different from
the one we started with as we demonstrated  in the previous example.

{\it (ii)}  On the other hand, if our gauge field arises from 
compactification of
the antisymmetric fields,we can have either $A^{(1)}_{\mu}$ or $A^{(2)}_{\mu}$
as we like, then both the dilaton and the gauge field will transform
to generate new background configurations.

It is important to note that the action considered in [31] 
has axion, dilaton, gauge field in addition to graviton. This 
action  is not invariant
under {\it their} $\bf {SL(2,R)}$ transformations ( ${\bf E}^2 - {\bf B}^2
\rightarrow {\bf B}^2 - {\bf E}^2$),
 under the duality transformation), 
however the equations of
motion are duality invariant.  
 The action considered by us has an axion from the RR sector
and the antisymmetric tensors $ H^{(i)}_{\mu \nu \rho}$ and 
gauge fields $ A^{(i)}_{\mu \alpha}$ transform
nontrivially under the S-duality transformation eq.(19). Furthermore, the action
itself is invariant under the symmetry transformation.   

Let us consider the following four dimensional effective action
\bea S_4 = \int d^4 x {\sqrt {-g}} [ R - {1\over 2} (\partial \phi)^2
- {1\over 4} e^{-\phi} F^{(1)}_{\mu \nu}F^{(1)\mu \nu}] \eea
The action of reference 31
can be obtained from the above one by scaling the dilaton
by a factor of two and removing the factor of ${1\over 4}$ from the gauge 
field strength squared term. We keep the superscript $1$ to remind that this
gauge field strength came from compactification of $H^{(1)}_{MNP}$.
In this case,  $\cal M$ is also diagonal in  absence of RR 
axion field, $\chi$.
 If we implement an $\bf {SL(2,R)}$ transformation, the new dilaton
and RR axion will be given by the same expression as eqns.(28)
and (29). However, now we
shall generate  new gauge field configurations
\bea A^{(1)'}_{\mu} = {1\over {\sqrt 2}}
(cosh{\theta} + sinh{\theta})A^{(1)}_{\mu}, \eea
\bea A^{(2)'}_{\mu} =  {1\over {\sqrt 2}}
(cosh{\psi} - sinh{\psi})A^{(1)}_{\mu} \eea
Recall that initially $A^{(2)}_{\mu} = 0$.  We mention in passing that if 
the four dimensional action had terms corresponding to $H^{(1)}_{\mu \nu \rho}$
and $H^{(2)}_{\mu \nu \rho}$ squares with $\cal M$ matrix, then a Poincare
duality transformation on these two field strengths would have given rise to
two additional 'axions'. We are considering a different scenario; however,
the action in eq.(34) does admit charged black hole solutions.

To summarize, the ten dimensional type IIB superstring 
action can be expressed in $\bf {SL(2,R)}$
invariant form with the introduction of the metric $\Sigma$. The compactified
theory on a d-dimensional torus respects the symmetry when we specify the
transformation properties of the resulting scalar and vector fields. It is
shown that the $\bf {SL(2,R)}$ invariant interactions terms involving only
$\cal M$-matrix results in adding a cosmological constant term even when we
 construct general form of 
invariants such as trace of the product $\Sigma \cal M$.
Since the action is invariant under the symmetry, we can construct new
backgrounds from a set of background which satisfies equations of motion.
We presented two illustrative examples: one in the case of five dimensional
black hole solution of Strominger and Vafa[13] and the other for the four 
dimensional black hole solutions of Shapere, Trivedi and Wilczek [31].

It is evident that 
$\bf {SL(2,R)}$ transformations together with the T-dualities
will enable us to generate a large family of gauge inequivalent  
background  confugurations when we implement them on type IIB vacuum
solutions in diverse dimensions.

\noindent
{\bf Acknowledgements:} I would like to thank Sumit R. Das and Sandip Trivedi
for very valuable discussions. 
I am grateful to the Jawaharlal Nehru Memorial Fund
for award of Jawaharlal Nehru Fellowship. The warm hospitality of Theory
Group of the Fermilab is gratefully acknowledged.

\end{document}